\documentclass[a4paper]{aa}

\usepackage{graphicx}
\usepackage{txfonts}
\usepackage{natbib}
\bibpunct{(}{)}{;}{a}{}{,}  

\begin{document}

   \title{Asteroseismology of Procyon A with SARG at TNG\footnote{Based on observations made
 with the Italian Telescopio Nazionale Galileo (TNG) operated on the island of 
La Palma by the Centro Galileo Galilei of the INAF (Istituto Nazionale di Astrofisica) 
at the Spanish Observatorio del Roque de los Muchachos of the Instituto de Astrofisica 
de Canarias}}

   \author{R. U. Claudi,
          \inst{1}
	  A. Bonanno,
	  \inst{2}
	  S. Leccia,
	  \inst{3}
	  R. Ventura,
	  \inst{2}
	  S. Desidera,
	  \inst{1}
	  R. Gratton,
	  \inst{1}
	  R. Cosentino,
	  \inst{2}	
	  L. Patern\`o
	  \inst{3}
          \and
          M. Endl\inst{4}
          }

   \offprints{R. U. Claudi}

   \date{Received ; accepted }

\institute{I.N.A.F. -- Oss. Astronomico di Padova, Vicolo Osservatorio 5, 35122
Padova, Italy\\
              \email{claudi@pd.astro.it,desidera@pd.astro.it,gratton@pd.astro.it}
         \and
            I.N.A.F. -- Oss. Astrofisico di Catania, Via S. Sofia 78, 95123 Catania, Italy
	    \\
             \email{abo@ct.astro.it,rve@ct.astro.it,rco@ct.astro.it}
	     \and
            Dipartimento di Fisica ed Astronomia, Universit\`a di Catania, Via S.
Sofia 78, 95123 Catania, Italy
	    \\
             \email{leccia@na.astro.it,lpaterno@ct.astro.it}
	     \and
           McDonald Observatory, The university of Texas at Austin, Austin, TX
78712, USA 
	    \\
             \email{mike@astro.as.texas.edu}
             }
\authorrunning{R.U. Claudi {\it et al.}}
\titlerunning{Asteroseismology of Procyon A with SARG}

   \abstract{
   We present high precision radial velocity measurements on the F5 IV star
$\alpha$ CMi obtained by the SARG spectrograph at TNG (Telescopio Nazionale Galileo)
exploiting the iodine  cell technique. The time series of about 950 spectra of Procyon A
 taken during 6 observation nights  are affected by an individual 
 error of  $1.3\,\rm m\,s^{-1}$. Thanks to the iodine cell technique, the
 spectrograph contribution to the Doppler shift measurement error is quite negligible and
 our error is dominated by the photon statistics \citep{brown94}.
 An excess of power between 0.5 and
 1.5 mHz, detected also by \citet{martic04} has been found. We determined a large separation
frequency $\Delta\nu_0 = 56\pm 2\,\mu\rm{Hz}$, consistent with both theoretical estimates
\citep{chaboyer99} and previous observations \citep{martic04}.

   }   
	\maketitle

\section{Introduction}
Procyon A ($\alpha$ CMi, HR 2943, HD61421) for its proximity and brightness has already
attracted attention of stellar seismologists \citep{guenther93, barban99,
chaboyer99, martic99}. It is a F5 IV star (with V$=0.363$) at a
distance of $3.53$~pc in a 40-year period visual binary system; the companion is
a white dwarf more than 10 mag fainter than the primary. Adopting the
very precisely measured parallax by \emph{Hipparcos}, $\Pi = 285.93\pm 0.88$  mas,
Prieto et al. (2002) derived a mass of 
$1.42 \pm 0.06$ $M_{\odot}$, a radius of
$R/R_{\odot} = 2.071 \pm 0.02$ and a gravity of $\log$ $\rm g = 3.96 \pm 0.02$. 

These parameters cannot unambigously determine the 
evolutionary status of the star, which could be either in the core 
hydrogen-burning phase or in the more advanced hydrogen shell-burning phase.
Although in principle seismology would help to establish the evolutionary status 
of Procyon A \citep{dimauro04}, the nature of its pulsational spectrum
is a subject of intense debate.

In fact, the excess of power around 1 mHz found by \citet{martic04} seems to have  a stellar origin and is consistent with a $p$-mode  comb-like pattern with
 a large frequency separation of about 54 $\mu$Hz.
However, data from the recently launched Canadian \emph{MOST} satellite \citep{
matthews04} show no significant power excess in the same spectral region.
While \citet{matthews04} suggested that the ground-based detection of  \emph{p}-modes may be an artefact
caused by a combination of stellar noise, and data sampling and analysis, \citet{
christensen04}
suggest that  the most likely explanation for the null
detection seems to be a dominating non-stellar noise source in the  \emph{MOST} data.

The aim of this letter is to discuss the observations of Procyon A carried out
using the high-resolution spectrograph SARG mounted at TNG (Telescopio Nazionale Galileo),
aimed at detecting stellar oscillations by means of high-precision radial velocity measurements.
Our findings confirm the presence of an excess of power around the 1 mHz region, 
where we detect several frequencies whose distribution is consistent with a $p$-mode
spectral pattern.
In addition, our results show  the high efficiency of the SARG spectrograph in detecting
solar-like oscillations in stars, being an ideal instrument for
multisite ground-based observing campaigns.

\section{Observations}
SARG is a high resolution cross dispersed echelle
spectrograph \citep{gratton01} which offers both single object and long slit (up to 26 arcsec)
observing modes covering a spectral range from $\lambda = 370\,\rm nm$ up to about 
$1000\,\rm nm$, with a resolution ranging from $R=29,000$ up to $R=164,000$.

Our spectra were obtained at $R=144,000$ in the wavelength range between 462 and 
$792\,\rm nm$. The iodine cell covers only the blue part (see Fig. 1) of the spectrum
(from 462 up to $620\,\rm nm$) which has been used for measuring stellar Doppler shifts. 
The red part of the echelle spectrum of Procyon A has been used for measuring equivalent width of absorption
lines sensitive to temperature. Here we present the results concerning the blue part of
the spectrum while the analysis of the red one will be the topic of a future work.
During the observations (the first SARG scientific run after commissioning and testing phases)
we collected about 950 spectra with high signal-to-noise ratio and a mean exposure time 
of about $10\,\rm s$. Due to weather  and technical conditions
(see Table 1 for more details) a few gaps are present in the time series.

\begin{figure}
   \centering
   \includegraphics[width=8cm]{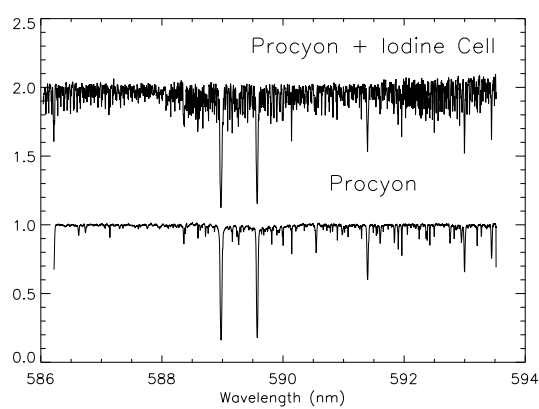}
      \caption{A segment of the echelle spectrum of Procyon obtained by SARG. Top panel: iodine 
absorption lines are superimposed to the Procyon lines. Bottom panel: the Procyon 
spectrum without iodine absorption features. This spectrum has been used as the
template spectrum in the determination of radial velocity from star plus iodine spectra}
         \label{Cella}
   \end{figure}

\begin{table}
\begin{center}
      \caption[]{Observing log: date, number of spectra, signal-to-noise ratio and seeing.}
        \begin{tabular}[h]{cccc} 
            \hline
             Date & Sp. number & SNR & Seeing (as) \\            
            \hline
           2001/01/02&160&325 $\pm$ 4&1.04 $\pm$ 0.02\\
           2001/01/03&149&248 $\pm$ 4&0.81 $\pm$ 0.01\\
           2001/01/04&181&292 $\pm$ 3&0.82 $\pm$ 0.01\\
           2001/01/06&181&295 $\pm$ 5&0.93 $\pm$ 0.01\\
           2001/01/08&125&291 $\pm$ 5&0.89 $\pm$ 0.02\\
           2001/01/09&153&328 $\pm$ 4&0.87 $\pm$ 0.01\\   
            \hline
	    \end{tabular}      
\end{center}
  \end{table}
   
\section{Data analysis}
Radial velocities have been obtained using the AUSTRAL code \citep{endl00} which
models instrumental profile, star and iodine cell spectra in order
to measure Doppler shifts.
Since the instrument profile changes along a spectral order, the spectrum was divided in 
segments (chunks) approximately 100 pixels long ($\sim$ 2 \AA), each one modeled independently. 
About 400 segments 
were used in the final analysis. The internal velocity error of a spectrum is calculated 
as the error of the mean velocity of all segments used for the analysis. The parameters 
for the instrument profile modelling were determined by using fast rotating featureless 
stars, the standard radial velocity star $\tau$ Cet and the planet-bearing  51 Peg 
and $\rho$ CrB stars. Detailed description of the technique is given in \citet{desidera03}.
Fig. 2 shows, as an example, the radial velocity time series obtained on 2001/01/06. 
In this case, the radial velocity internal error and the r.m.s. of the data set are  
$1.48\,\rm m\,s^{-1}$ and $3.5\,\rm m\,s^{-1}$, respectively. A 21-min pulsation pattern is clearly visible.
\begin{figure}
   \centering
   \includegraphics[width=8cm]{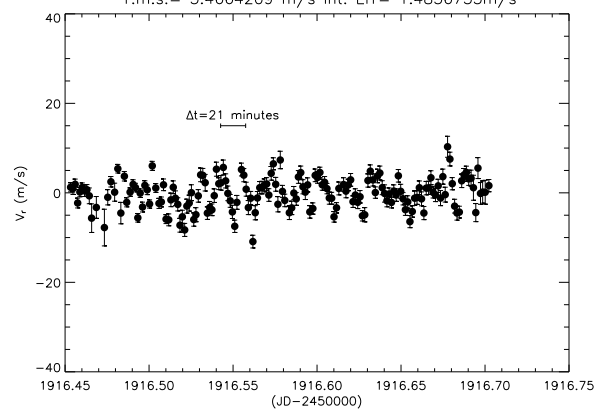}
      \caption{Radial velocities of Procyon obtained on 2001/01/06. The data are affected by
an  internal error of $1.48\,\rm m\,s^{-1}$ and an r.m.s. of $3.5\,\rm m\,s^{-1}$.
The 21-minute pulsation is also reported.}
         \label{Cella}
   \end{figure}

   \begin{figure}
\centering
\includegraphics[width=8cm]{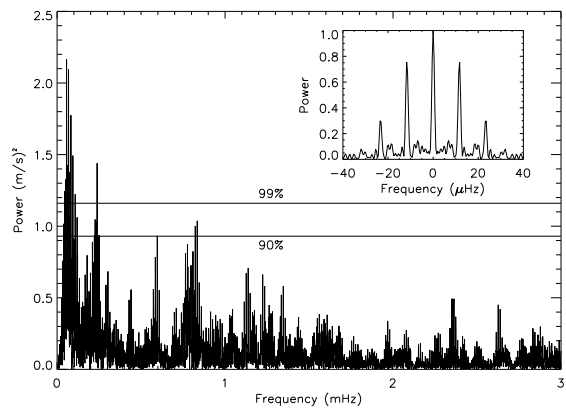}
\caption{The Scargle-Lomb periodogram of the data. An excess of power around 1
mHz is evident. The horizontal solid lines show the confidence levels
corresponding to the probability of detecting a peak due to a genuine signal
equal to 90\% and 99\%. The inset shows the power spectrum of the window
function for a sine wave signal of amplitude 1 m s$^{-1}$, sampled as the
observations. The power unit are the same as the main figure.}
\label{fig:period}
\end{figure}

We have used the modified Scargle-Lomb periodogram \citep{scargle82} in order to perform the spectral analysis.
The result obtained  by combining all the six observing nights is shown in Fig. 3, where an excess of power around 1 mHz is clearly visible. 
The time scale gives a spectral resolution of 1.62 $\mu$Hz.
The calibration of the Scargle's power in terms of amplitude squared was performed in such a way that a synthetic sine wave of 1 m s$^{-1}$, sampled as the observations,
would yield in the periodogram a cluster of peaks, the largest of which with a power equal to 1 m$^2$ s$^{-2}$.
We then computed the statistical significance of the detected peaks in terms of the "false alarm probability" F, 
as defined by \citet{scargle82}, and adopting the prescription of \citet{horne86}. F is
the probability that a spike (i.e. power in one frequency bin) is due to noise. 
Consequently the quantity (1-F) is the probability of detecting a peak due to a genuine signal. 
The power levels (confidence levels) concorresponding to the value (1-F) equal to 90\% and 99\%, 
equivalent to a false alarm probability of 10\% and 1\%, respectively, are reported in Fig. 3.
In Table 2 we listed all the individual frequencies detected with
a confidence level greater than 90\%; the frequencies consistent,  with those found by \citet{martic04} are shown in bold-face.
The additional frequencies of 302.0 $\mu$Hz, 428.0 $\mu$Hz and 808.8 $\mu$Hz
are also consistent with those of \citet{martic04}, but have been detected with
a confidence level smaller than 90\%.
By comparing the power spectrum shown in Fig. 3 with the results presented in
\citet{martic99}, it is worth noting that most of the strongest peaks in our
periodogram have power of about 0.9 m$^2$ s$^{-2}$, a value greater than that
found by \citet{martic99} in the analysis of the observations carried out on
November 1997 using the ELODIE spectrograph, which ranges from 0.8 m$^2$ s$^{-2}
$ in the periodograms obtained for two subsets of data to 0.5 m$^2$ s$^{-2}$ in
the periodogram of the entire data set. On the other hand the high frequency
mean white noise level in our periodogram, evaluated in the frequency range
 1.8 - 3 mHz, is 0.049 m$^2$ s$^{-2}$, while the values reported by \citet{martic99}  range from 0.014
to 0.030 m$^2$ s$^{-2}$, depending on the different set of data analyzed.
The power of most of the strongest peaks reported by \citet{martic04} referring to the ELODIE 
and the combined ELODIE/AFOE data sequences colleced on January-Febraury 1999 
is about 0.16 - 0.20 m$^2$ s$^{-2}$, while that reported for the AFOE data is 
about $0.5 - 0.6\,{\rm m^2\,s^{-2}}$; the mean noise levels are quite different
in the three
sets of data, ranging from 0.0256 to 0.0064 m$^2$ s$^{-2}$.
\begin{table}
\caption[]{ 
Prominent peaks present in the periodogram of Procyon A
detected by SARG with a
confidence level larger than 90\%: bold
characters indicate the frequencies consistent, within our spectral resolution
of 1.62 $\mu$Hz, with those found by \citet{martic04} with a spectral resolution of 
1.79 $\mu$Hz.}
\begin{center}
\begin{tabular}{ccccccc}
\hline
\multicolumn{7}{c}{Frequencies ($\mu$Hz)} \\
\hline
230.2 & 233.7 & 238.4 & 245.4 & {\bf 596.5} & {\bf 776.7} & {\bf 822.8}\\
\hline
\end{tabular}
\end{center}
\end{table}

In order to estimate the large frequency separation $\Delta\nu_0$ in the region of the
excess of power detected in the periodogram, we CLEAN-ed the spectrum by the window
function \citep[for the details of the adopetd procedure see][]{roberts87} and  successively
we applied to the CLEAN-ed power spectrum the comb-response (CR) method \citep{kjeldsen95}
which is a generalization of the power spectrum of a power spectrum
and consequently allows us to search for any regularity in a spectral pattern. 
In the case of solar-like oscillations it allows us to detect their typical comb-like structure.
In particular, a peak in the CR at a particular spacing $\Delta\nu_0$ indicates
the presence of a regular series of peaks in the power spectrum, centered at the
central frequency $\nu_0$
and having a spacing of  $\Delta\nu_0/2$. We adopted
the modified comb-response function \citep{martic04}:
   \begin{equation}
   \begin{array}{l}
      C(\nu_0,\Delta\nu_0)=\prod_{i=0}^4\lbrack
S(\nu_0+i\frac{\Delta\nu_0}{2}+\delta)S(\nu_0-i\frac{\Delta\nu_0}{2}+\delta)\rbrack
^{\alpha_i}
   \end{array}
   \end{equation}
where $\delta=\pm(i$ mod $2)D_0$ according to mode degree $l=0$ or $l=1$ at the
central frequency $\nu_0$ and $\alpha_i=1$ for $i=1$ or $2$, $\alpha_i=0.5$ for
$i=3$ or $4$.
   \begin{figure}
\centering
\includegraphics[width=8cm]{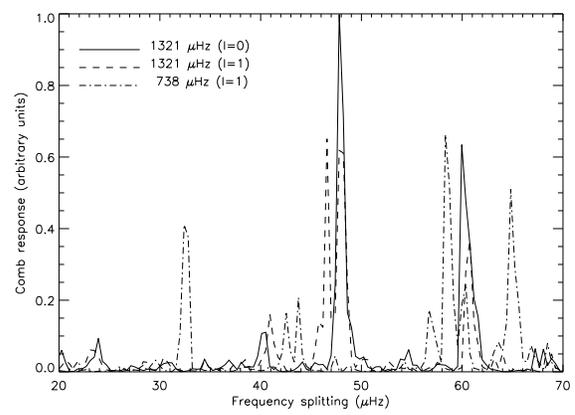}
\caption{{\bf Comparison between modified comb responses for the two central frequencies 
$738\,{\rm\mu Hz}$ and $1321\,{\rm\mu Hz}$ of the
power spectrum of Procyon A data. The responses are scaled to the maximum peak obtained 
for the frequency $\nu_0 = 1321\,{\rm \mu Hz}$ at $\Delta\nu_0 = 47.8\,{\rm \mu Hz}$.}}
\label{fig:comb}
\end{figure}
   \begin{figure}
\centering
\includegraphics[width=8cm]{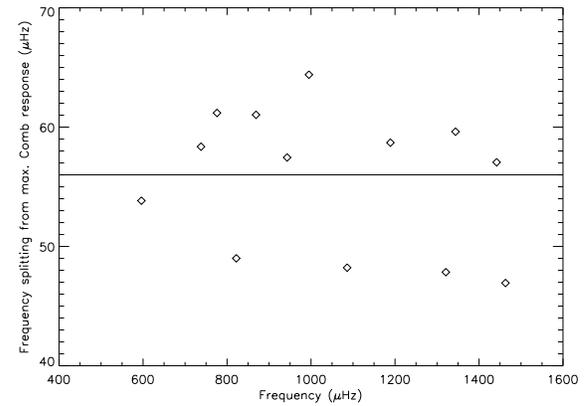}
\caption{{\bf The first-order spacings over several frequency ranges, as
determined from comb analysis. The line shows the mean value of
$56\,{\rm \mu Hz}$.}}
\label{fig:mean}
\end{figure}
For each central frequency $\nu_{0}$, alternatively considered as $l=0$ and $l=1$,
we searched for the maximum CR in the intervals $20\le \Delta\nu_0 \le 80$
 $\mu Hz$ and $0.3 \le D_0 \le 2$.
Fig. \ref{fig:comb} shows examples of the modified CR computed for the two
central frequencies 738 $\mu$Hz and 1321 $\mu$Hz.
In Fig. \ref{fig:mean} we show the variation of $\Delta\nu_0$ in the frequency range
$500 \leq \nu_{0} \leq  1500\,{\rm \mu Hz}$, as determined from CR analysis.
The arithmetic mean of this spacing  is $56\pm 2\,\mu\rm{Hz}$, a value of
the large separation in good agreement with both theoretical estimates
\citep{chaboyer99} and previous observations \citep{martic04}. The error on
$\Delta\nu_0$ has been estimated as the standard deviation of the
$\Delta\nu_0$'s computed from CR analysis.

The determination of $\Delta\nu_0$ is not strongly sensitive to the  choice of $D_0$.

\section{Conclusions}
The analysis of the periodogram of Procyon A obtained by SARG shows a $p$-mode
spectrum characterized by a large frequency separation $\Delta\nu_0=56\pm 2\,\mu\rm{Hz}$ 
in agreement with \citet{chaboyer99} and \citet{martic04}.
Contrarily to what found by \citet{matthews04} our results strongly support the
idea that the excess of power detected by \citet{martic04} in the range from 0.5 up to 
$1.5\,\rm{mHz}$ 
is caused by \emph{p}-mode oscillations. 
In particular we think that the SARG spectrograph can efficiently be used in future multi-site
observing campaigns for asteroseismology.

\begin{acknowledgements}
We wish to thank H. Kjeldsen and J. Christensen-Dalsgaard for useful comments
and suggestions.The authors wish to thank the anonymous referee for useful comments and
suggestions which contributed to improve the paper.

This work has partially been supported by MIUR (Italian Ministry of Education, University and
Scientific Research) under contract COFIN 2002, No 2002024732.
\end{acknowledgements}

\bibliographystyle{aa}

\end{document}